\documentclass[a4paper,10pt]{article}
\usepackage[utf8x]{inputenc}
\usepackage{amsmath,amssymb,graphicx,bm}
\usepackage{color}
\usepackage[bottom]{footmisc}
\newcommand{\beq}{\begin{equation}}
\newcommand{\eeq}{\end{equation}}
\newcommand{\bea}{\begin{eqnarray}}
\newcommand{\eea}{\end{eqnarray}}
\newcommand{\ba}{\begin{align}}
\newcommand{\ea}{\end{align}}
\newcommand{\bfig}{\begin{figure}}
\newcommand{\efig}{\end{figure}}

\begin{document}

\title{The 27-plet contributions to the $CP$-conserving $K\rightarrow \pi l^+l^-$ decays}

\author{B.\ Ananthanarayan, I. Sentitemsu Imsong \\
Centre for High Energy Physics, \\
Indian Institute of Science, Bangalore 560 012, India }



\maketitle

\begin{abstract}
We revisit the the rare kaon decays $K\to \pi l^+l^-$ which are of 
interest specially due to the recent measurements of the charged kaon decay 
spectra.  We compute the contribution of the 27-plet to
the decay amplitudes in one loop $SU(3)$ Chiral Perturbation
Theory.  We estimate the resulting impact to be $\sim 10\%$
to the branching ratios of the charged kaon decays, and also
noticeably influence the shape of the spectra.  With current values of
the constants $G_8$ associated with the octet and $G_{27}$
associated with the 27-plet, the contribution of the latter
pushes the spectrum in the correct direction for the charged
lepton spectra.  We also discuss the impact for the neutral
decay rates and spectra.
\end{abstract}

\bigskip


Kaon decays are of great importance to test the consistency of the standard
model at low energies.  Semi-leptonic and non-leptonic decays have been 
studied extensively within the framework of one loop SU(3) chiral 
perturbation theory \cite{Gasser:1984gg,Gasser:1984ux}, 
which is the effective low-energy theory of the Standard Model 
involving the pseudo-scalar mesons. 
These have been recently reviewed in great 
detail in ref.\cite{Cirigliano:2011ny}.
Of special interest is the charged
kaon decay to a pion and a charged lepton pair which has been recently studied 
at high precision by the $NA48$ collaboration \cite{Batley:2009pv,Batley:2011zz}. 
The processes we are considering in this paper include
the decays above, and are the $CP$-conserving 
decays $K^+\rightarrow \pi^+ l^+l^-$ and $K_S\rightarrow \pi^0 l^+l^-$ $(l=e,\mu)$.
The charged decays corresponding to the electron and muon channels {\it viz}, $K^+\rightarrow \pi^+ e^+e^-$ and 
$K^+\rightarrow \pi^+ \mu^+\mu^-$ have been analysed in
refs. \cite{Batley:2009pv} and \cite{Batley:2011zz} respectively.
In addition to studying the branching ratios, the collaboration 
has also gathered data on the decay lepton spectrum.  
The analysis of the data has rested on the classic treatment 
of ref.\cite{Ecker:1987qi} and its extensions.
The basis of the computation in ref.~\cite{Ecker:1987qi} is one loop $SU(3)$
Chiral Perturbation Theory (ChPT) for non-leptonic decays, and these 
have also been reviewed in ref.\cite{D'Ambrosio:1994ae}
(see also, ref.~\cite{Cirigliano:2011ny}). 
In order to go beyond $O(p^4)$, unitarity corrections are accounted using a
dispersive treatment, see ref. \cite{D'Ambrosio:1998yj}. 
Further, some works have been done for isospin breaking
effects and radiative corrections of these decays in \cite{Bijnens:2007xa}
and \cite{Kubis:2010mp} respectively.
The processes have also been discussed in interesting theoretical scenarios such as one in which
ChPT is combined with Large-$N_c$ QCD~\cite{Friot:2004yr}.
The experimental analysis in ref.~\cite{Batley:2009pv,Batley:2011zz}
is based on fitting the data to various models involving form factor parameterisations
to extract the relevant parameters of the models, which is inspired by
all the extensions described above.  It may also be noted that
the processes have also been considered in the context of so-called weak static 
interactions ~\cite{Dubnickova:2006mk}. 

Considerably less information is available in the case of the neutral
kaon decays.  It is also known that the one loop ChPT computations lead
to results that are very sensitive to the values of theoretical inputs.
For the sake of completeness we also present the amplitude for these
decays and provide a numerical treatment for the decay spectra.

Each of the processes of interest is assumed to be dominated by the 
one virtual photon exchange contribution $(K\rightarrow \pi \gamma^*)$ 
and that it is dominated by the octet part of
the non-leptonic effective Lagrangian characterized by the coupling 
constant $G_8$.
We note here that there is a sub-dominant piece in the effective Lagrangian
that is associated with a piece that transforms as 27-plet of
$SU(3)$ flavour denoted by $G_{27}$.  The 27-plet has been expected
to be sub-dominant as $G_{27}\simeq G_{8}/18$~\cite{D'Ambrosio:1994ae}.
The associated Lagrangian can be found in
\cite{Cronin:1967jq,Kambor:1989tz,Ecker:1992de} 
and can be expressed in short-hand as - 
\begin{equation}
 {\cal L}_4 =  {\cal L}_4^8 +  {\cal L}_4^{27}
\end{equation}

The corresponding Low-Energy Constants (LEC's) associated with these Lagrangians are denoted by $N_i$ and $R_i$
\footnote{The constant $R_i$
is also denoted by $D_i$ in the literature, see for instance \cite{Bijnens:2004ku,Bijnens:2004vz,Bijnens:2004ai}.}  
also enter the amplitudes at one-loop chiral perturbation theory, which are associated with the octet
and the 27-plet respectively. The contribution of the 27-plet has been accounted for in some processes,
such as $K_L \rightarrow \pi \gamma \gamma$ \cite{Cappiello:1988yg,Cappiello:1992kk}, and the result for the amplitude was
recently verified by Unterdorfer and Ecker (UE) 
in ref.\cite{Unterdorfer:2005au} where they have also 
partly studied the process $K_S\rightarrow \pi^0 \gamma \gamma$.

In the present work, we turn our attention to the contributions of the
27-plet to the process $K \rightarrow \pi l^+ l^-$ in one-loop $SU(3)$ ChPT. 
Our work is motivated by the need to provide completeness to the treatment of
this amplitude. The one loop result of ref. \cite{Ecker:1987qi} has attracted considerable attention, see for
instance ref. \cite{D'Ambrosio:1994ae} where a detailed consideration of the phenomenological consequences was taken
up. It has been observed that the neutral kaon rate is very sensitive to the choice of
the LEC's which makes a precise prediction virtually impossible.
On the other hand, there is considerable uncertainty in the value of $G_8$ which has come
down quite significantly due to constraints from charged kaon branching ratios. 
Using more recent values of $G_8$ and also including the uncertainties allows us to
assess the impact of the contributions due to the 27-plet piece of the amplitude.

The coefficients $G_8$ and $G_{27}$ have
been discussed in several work, for instance see 
refs. \cite{Ecker:1987qi,D'Ambrosio:1994ae,D'Ambrosio:1996zy,D'Ambrosio:1997tb,Bijnens:2004ku,Bijnens:2004vz,Bijnens:2004ai,Cirigliano:2003gt}. 
It is therefore clear that this process is considered as a very important test of the consistency 
of low-energy effective theories and hence merits further attention. 

We use the standard definitions of the Lagrangian presented in ref.\cite{Unterdorfer:2005au} (also consistent with the
conventions in ref.\cite{Cirigliano:2011ny}). We employ the software made available by UE and
described in great detail in ref. \cite{Unterdorfer:2005au} which readily allows us 
to compute the amplitudes for the decays 
$K \rightarrow \pi \gamma^*$ for the charged as well as the neutral cases. From general considerations, 
it may be recalled that we have the general structure for this amplitude which is given by -
\begin{equation}
A(s) = -\frac{G_F \alpha}{4\pi}V_i(s)(k+p)^\mu \bar{u_l}(p_-)\gamma_\mu v_l(p_-), \,\,\,\,\,\, i = +, S
\end{equation}

 The amplitudes for the two cases of interest read:
\begin{multline}\label{eq:charged}
V_+(s) = \frac{1}{54G_Fs}[3G_8[48\pi^2(s-4M_K^2)\bar{J}(s,M_K^2) +  
48\pi^2(s-4M_\pi^2) \bar{J}(s,M_\pi^2) \\
- s[3 \, {\rm ln}(\frac{M_K^2}{\mu^2}) - 384\pi^2 (3L_9+N_{14}-N_{15}) + 3 \,{\rm ln}(\frac{M_K^2}{\mu^2}) + 2 ]] 
+ G_{27}[-624\pi^2 \\
(s-4M_K^2)\bar{J}(s,M_K^2)- 624\pi^2(s-4M_\pi^2) \bar{J}(s,M_\pi^2) + 
s[13[3 \,{\rm ln}(\frac{M_K^2}{\mu^2})+ 3 \,{\rm ln}(\frac{M_\pi^2}{\mu^2}) \\
+2]+ 576\pi^2 (4L_9- R_{13}+2R_{15})]]]
\end{multline}

\begin{multline}\label{eq:neutral}
V_S(s) = \frac{1}{54G_Fs(M_K^2-M_\pi^2)}[G_{27}[48\pi^2(6 M_\pi^2 - M_K^2)(s-4M_K^2)\bar{J}(s,M_K^2) \\
+ 240\pi^2(2 M_\pi^2- 3 M_K^2)(s-4M_\pi^2) \bar{J}(s,M_\pi^2) + M_K^2s[3 \,{\rm ln}(\frac{M_K^2}{\mu^2})+ 45 \,{\rm ln}(\frac{M_\pi^2}{\mu^2}) - 
\\
576\pi^2R_{13}+16]] - 2M_\pi^2s[9 \,{\rm ln}(\frac{M_K^2}{\mu^2})+ 
15 \,{\rm ln}(\frac{M_\pi^2}{\mu^2}) -288\pi^2R_{13}+8]]- 6G_8 \\
(M_K^2-M_\pi^2)[s[3 \,{\rm ln}(\frac{M_K^2}{\mu^2})-
96\pi^2(2N_{14}+N_{15})+1]-48\pi^2(s-4M_K^2)\bar{J}(s,M_K^2)]]
\end{multline}
respectively for the charged and neutral decays, where
\bea
\bar{J}(s, M^2) &=& \frac{1}{16\pi^2}[\sigma(s,M^2)\, {\rm ln}\frac{\sigma(s,M^2) - 1}{\sigma(s,M^2) + 1} + 2] \\
\sigma(s,M^2) &=& \sqrt{1- \frac{4M^2}{s}} 
\eea

We note that the loop contribution to the charged case of the 
$G_{27}$ has the same
algebraic structure as that of $G_{8}$ and differs only by an overall factor.
In the neutral case, we have $G_{27}$ contributions involving both kaon and
pion loops which do not have the same algebraic structure as that of
the $G_{8}$ loop contributions which involves only the kaon loops.
We have checked that the part of the amplitude when
we put $G_{27}$ to zero, agrees with the amplitudes given in
\cite{Cirigliano:2011ny,D'Ambrosio:1994ae}. 

One may express the differential rate in terms of a dimensionless variable 
$z = s/M_K^2$, and is   given by -
\begin{equation}\label{eq:diffdecay}
\frac{d\Gamma}{dz} = \frac{G_F^2\alpha^2M_K^5}{12\pi(4\pi)^4}\bar{\lambda}(1, z, r_\pi^2)^{\frac{3}{2}}
\sqrt{1-\frac{4r_l^2}{z}}(1+\frac{2r_l^2}{z})|V_{+,S}(z)|^2
\end{equation}
with
\begin{equation}
4r_l^2 \leq z\leq (1-r_\pi)^2 \nonumber
\end{equation}
and where,
\bea \label{eq:defn}
\bar{\lambda}(a, b, c) &=& a^2 + b^2 + c^2 - 2(ab + bc + ca) \\  
r_l &=& \frac{M_l}{M_K} \\ 
r_\pi &=& \frac{M_\pi}{M_K}
\eea

It is customary to discuss the values of $G_8$ and $G_{27}$ in terms of associated constants 
denoted by $g_8$ and $g_{27}$ which are related by - 
\begin{equation} \label{eq:convert}
G_{8,27} = -\frac{G_F}{\sqrt{2}}V_{ud}V_{us}^*g_{8,27}
\end{equation}

We collect some information on $g_8$ and $g_{27}$ from the literature.
A fit from $K \rightarrow 2\pi$ decays yields $g_{8} = 5.1$ \cite{Ecker:1987qi}
while more recent evaluations \cite{Cirigliano:2011ny} at next-to-leading order with isospin violating
effects included from a fit to $K \rightarrow 2\pi$ decays yields $g_{8} = 3.61 \pm 0.002_{\rm exp} \pm 0.28_{\rm th}$ 
and $g_{27} = 0.297 \pm 0.0006_{\rm exp} \pm 0.028_{\rm th}$.
A fit of both the $K \rightarrow 2\pi$ and  $K \rightarrow 3\pi$ amplitude calculated in ChPT at 
NLO \cite{Bijnens:2004vz} with various available data for $K \rightarrow 2\pi, 3\pi$ decays
yields $g_{8} = 3.27$ and $g_{27} = 0.235$.
On the other hand, the knowledge of the higher order LEC's namely the constants $N_i$ and $R_i$ is limited. 
Analysis of these constants and their estimates are given in several papers, see for instance 
\cite{D'Ambrosio:1996zy,D'Ambrosio:1997tb,Bijnens:2004ku,Bijnens:2004vz,Cirigliano:2003gt}.
For the purpose of illustration, we use in our analysis the values for $g_8$
and $g_{27}$ given in \cite{Cirigliano:2011ny} (quoted above) and 
the various constants given in \cite{Bijnens:2004vz} where, $L_9 = 7 \times 10^{-3}$,
$N_{14} = -10.4 \times 10^{-3}$, $N_{15} = 5.95 \times 10^{-3}$, $R_{13} = 0$ and
$R_{15} = 0$. The values of $G_8$ and $G_{27}$ corresponding to $g_8$ and
$g_{27}$ can be obtained using the expression given in eqn.(\ref{eq:convert}).
It may be noted that the renormalization scale is taken to be $\mu = M_\rho$.

It may be noted that apart from the values of the LEC's and the octet and 27-plet constants, there are
no other inputs in these theoretically clean profiles that appear
in these figures. In other words, in the present work we have
not attempted to include possible higher order O($p^6$) effects
of the type considered in the absence of the 27-plet in ref.~\cite{D'Ambrosio:1998yj}
which can considerably alter the shape of these profiles.
We will present a discussion on thsi as well.

We present below the results of our numerical analysis for the spectra
where we give the figures for the differential spectra for the 
$K^+ \rightarrow \pi^+l^+l^-$ in Fig. \ref{fig:fig1} and
Fig. \ref{fig:fig2} for the electron and muon
respectively. In Fig. \ref{fig:fig3} and Fig. \ref{fig:fig4}
we show the corresponding spectrum for the neutral decays $K_S \rightarrow \pi^0l^+l^-$.
In all these figures, we show the contribution in the absence of the 27-plet
contributions and in the presence of the same. We have shown a set of profiles
corresponding to the values of $G_8 (g_8)$ varied within errors and compared with 
the curve obtained by including the $G_{27}(g_{27})$ part.
Here, we have fixed $G_{27}$ at the central value and varied $G_8$ within the quoted errors
as mentioned. A crucial observation in Figs. \ref{fig:fig1} and \ref{fig:fig2} is that 
of the curve corresponding to the higher value of
$G_8$ which is comparable to the total curve {i.e,} including both the $G_8$ and $G_{27}$. 
In fact, the latter lies slightly below the the curve obtained with the 
higher value of $G_8$ alone which shows that the errors attached with the 
dominant octet piece yields the same effect compared
to the total amplitude with using just the central values of inputs. This behaviour is not seen in the neutral case and
that all the curves lie below the curve corresponding to the full amplitude. 
In the figures corresponding to the charged kaon decays, we have also
overlaid the data from the recent experiments, ref.~\cite{Batley:2009pv,Batley:2011zz},
which we have read off from the figures in those papers.
It may be seen that the clean figures that we have provided do not
adequately describe the experimental information. In both the electron and the muon case,
there is a crossover where the curves are higher for lower values
of $z$ than the experimental data, and {\it vice versa} for larger
values of $z$, see Figs. \ref{fig:fig1} and \ref{fig:fig2}.
It has been pointed out that the curves are very sensitive to the values
of the LEC's.  Thus a combined analysis including the contributions
of the 27-plet to the amplitude and unitarity corrections could be necessary.

\bfig[ht]	
	\begin{center}\vspace{0.cm}
	 \includegraphics[angle = 0, clip = true, width = 2.8 in]
{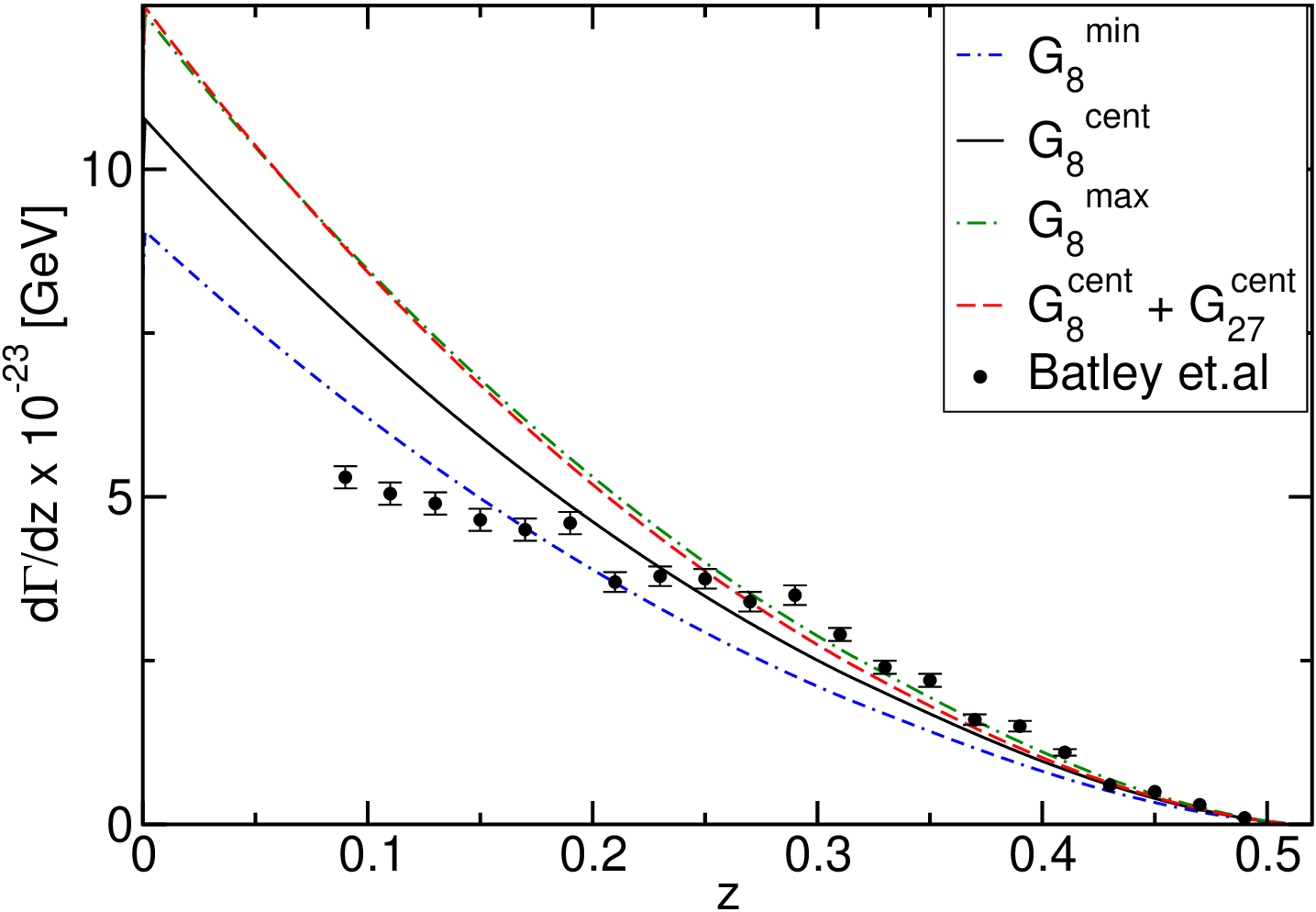}	
	\end{center}\vspace{-0.2cm}
	\caption{Spectrum for $K^+ \rightarrow \pi^+e^+e^-$ 
in the dilepton invariant mass in the presence and absence of $G_{27}$ where 
the $G_8$ part have been varied within errors. 
Also shown are the data extracted from the figures in \cite{Batley:2009pv}.} 
	\label{fig:fig1}
\efig	

\bfig[ht]	
	\begin{center}\vspace{0.cm}
	 \includegraphics[angle = 0, clip = true, width = 2.8 in]
{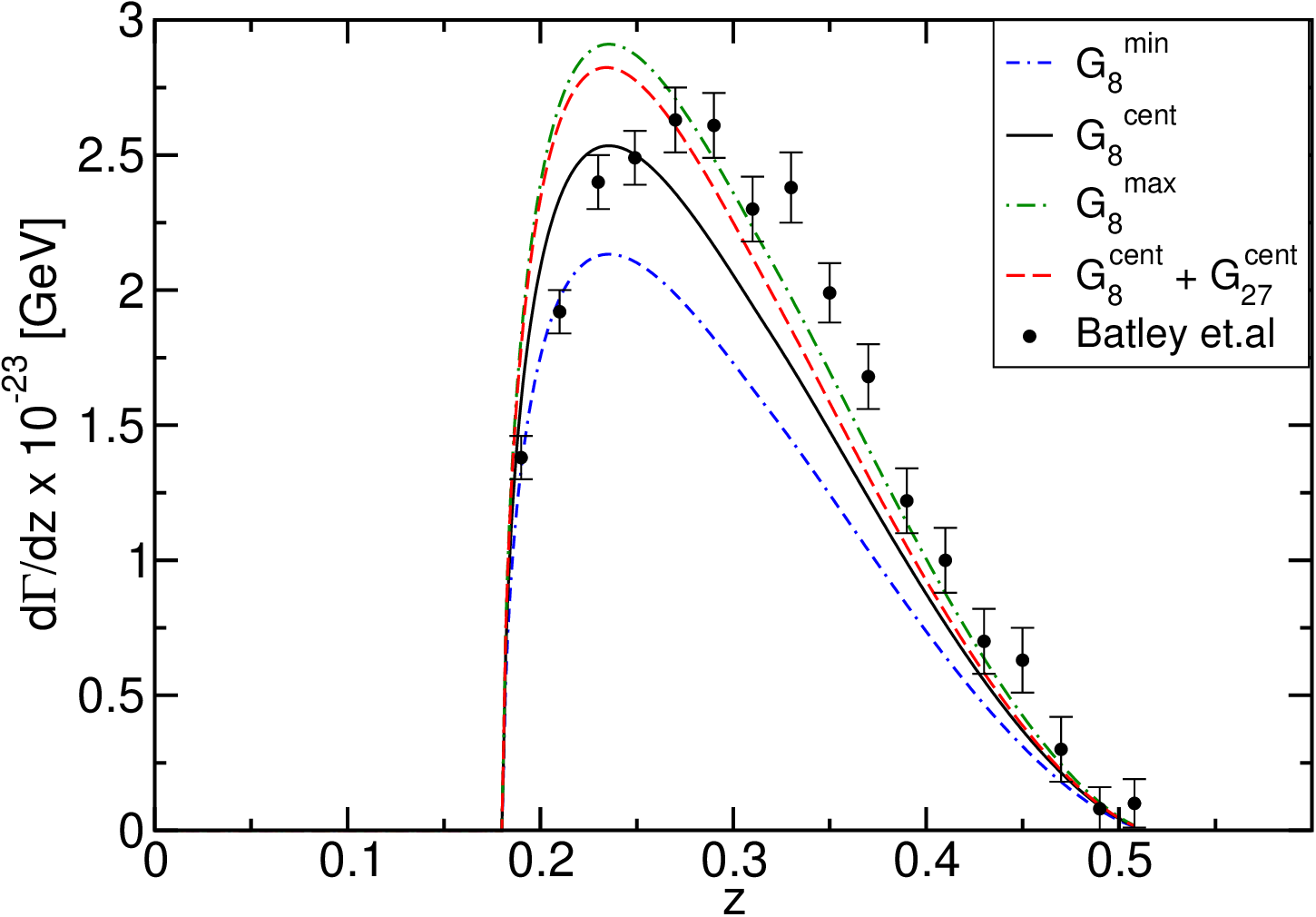}	
	\end{center}\vspace{-0.2cm}
	\caption{Spectrum for $K^+ \rightarrow \pi^+\mu^+\mu^-$             
in the dilepton invariant mass in the presence and absence of $G_{27}$
where the $G_8$ part have been varied within errors.  
Also shown are the data extraced from the figures in \cite{Batley:2011zz}.}  
	\label{fig:fig2}
\efig

\bfig[ht]	
	\begin{center}\vspace{0.cm}
	 \includegraphics[angle = 0, clip = true, width = 2.8 in]
{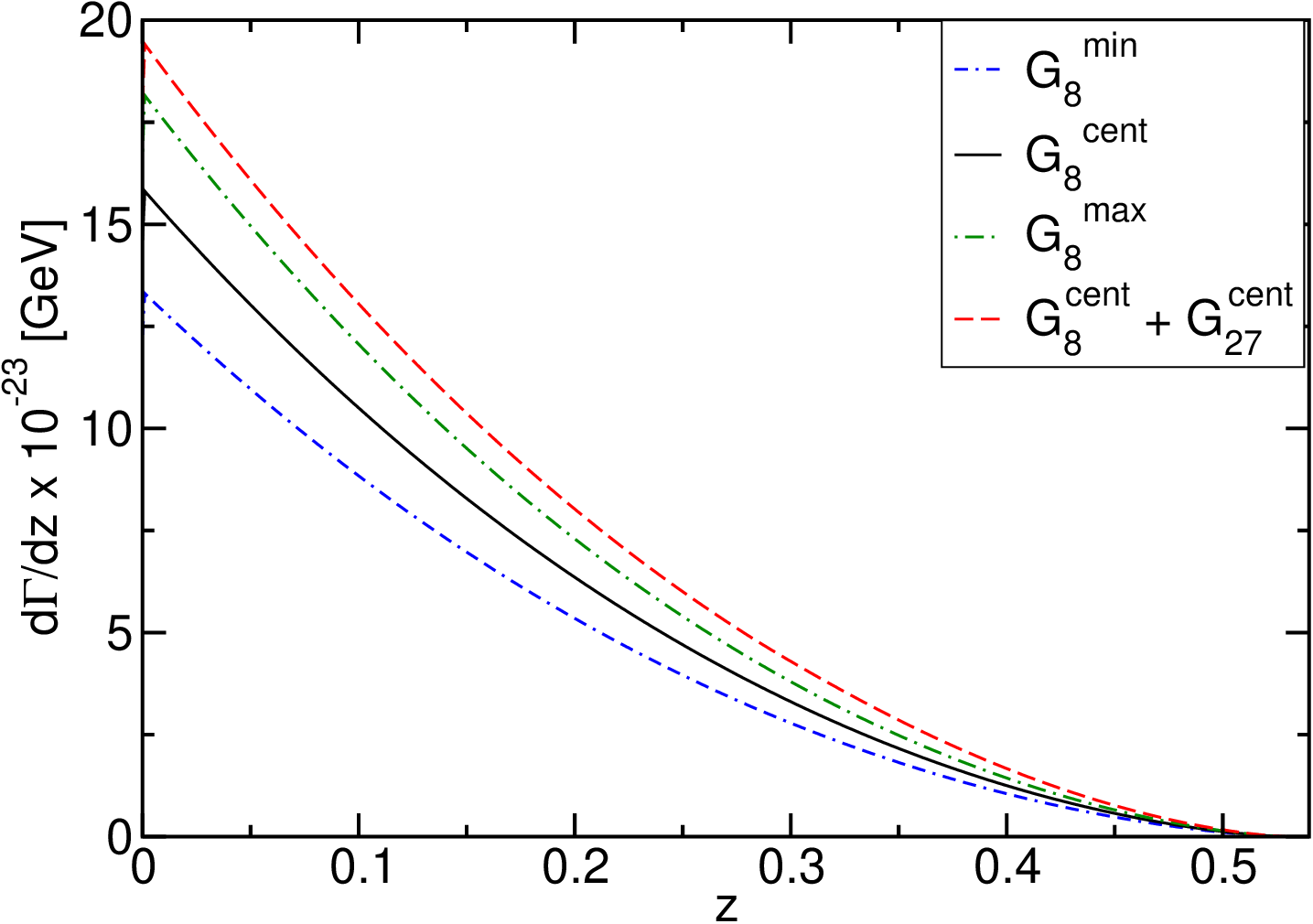}	
	\end{center}\vspace{-0.2cm}
	\caption{Spectrum for $K_S \rightarrow \pi^0e^+e^-$ 
in the dilepton invariant mass in the presence and absence of $G_{27}$ where 
the $G_8$ part have been varied within errors. .}  
	\label{fig:fig3}
\efig	

\bfig[ht]	
	\begin{center}\vspace{0.cm}
	 \includegraphics[angle = 0, clip = true, width = 2.8 in]
{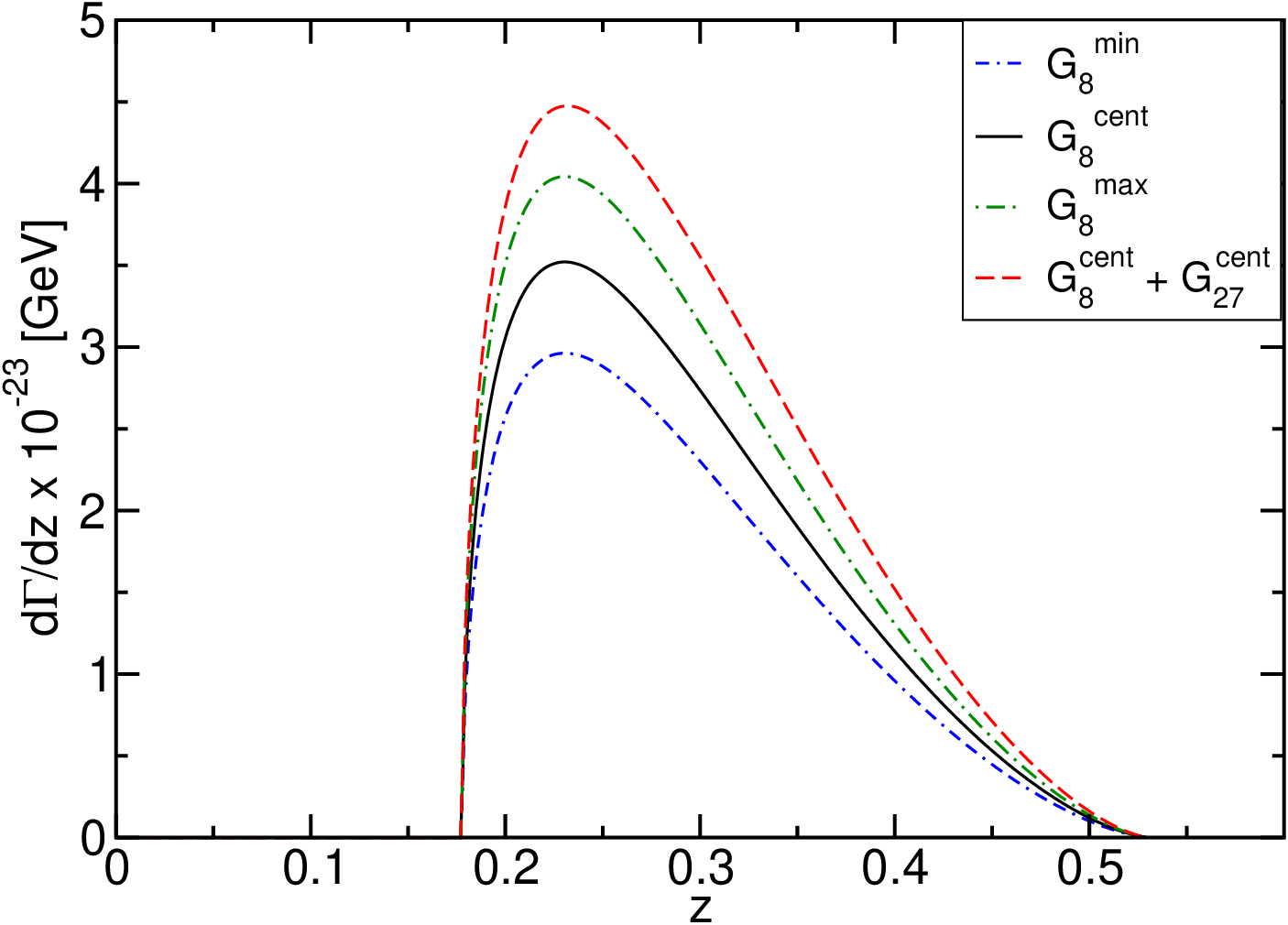}	
	\end{center}\vspace{-0.2cm}
	\caption{Spectrum for $K_S \rightarrow \pi^0\mu^+\mu^-$ (right)
in the dilepton invariant mass in the presence and absence of $G_{27}$ where 
the $G_8$ part have been varied within errors.}  
	\label{fig:fig4}
\efig

\bfig[ht]	
	\begin{center}\vspace{0.cm}
	 \includegraphics[angle = 0, clip = true, width = 2.8 in]
{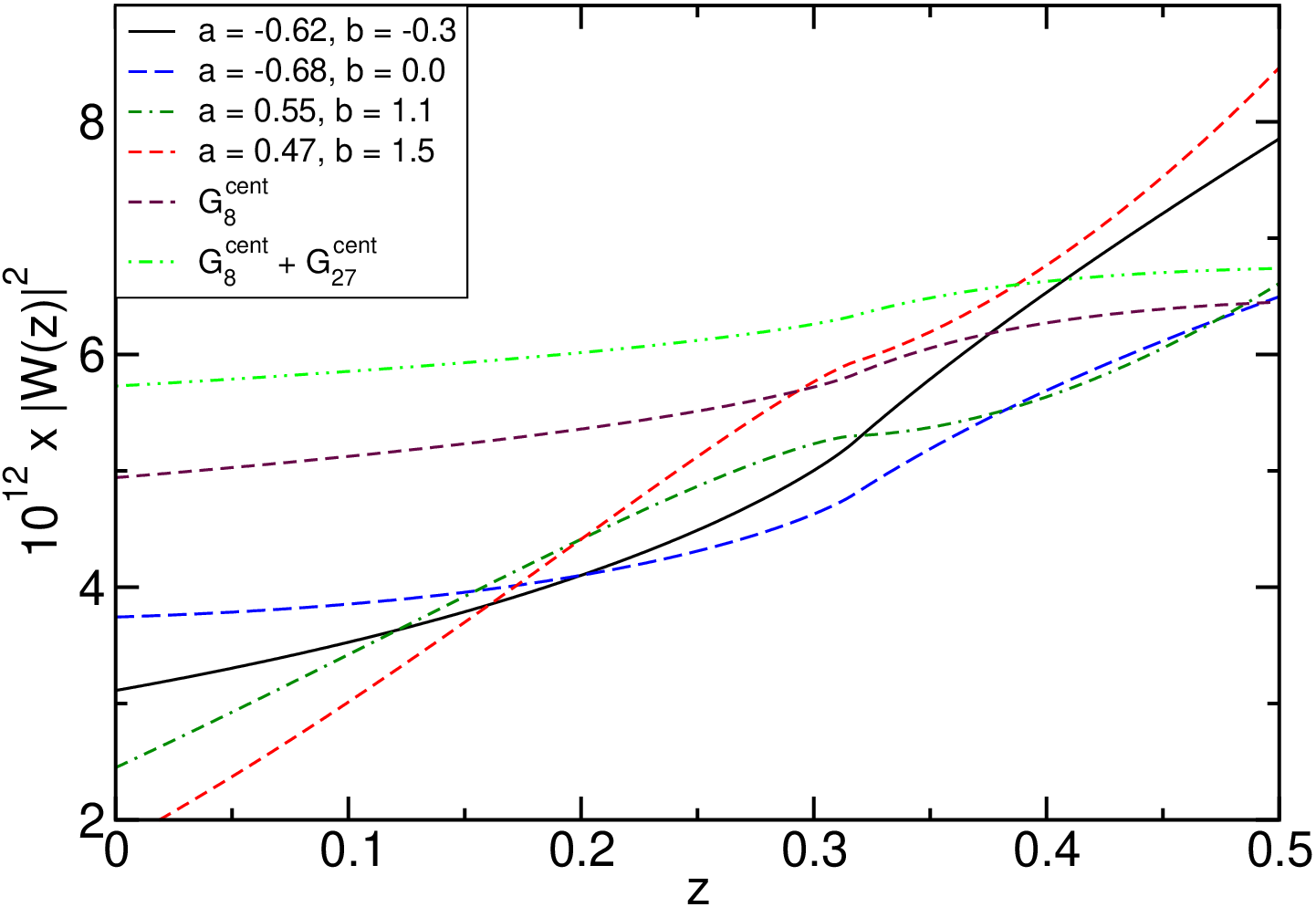}	
	\end{center}\vspace{-0.2cm}
	\caption{Plot of the quantity $W(z)$ as a function of $z$ for the process $K^+ \rightarrow \pi^+l^+l^-$. 
Here we give our results in the presence and absence of $G_{27}$
and compare with the unitarity corrections given in \cite{D'Ambrosio:1998yj} for different sets of parameters $a$ and $b$. }  
	\label{fig:fig5}
\efig

The branching ratios of each of these decays corresponding to the central
values of the inputs used are presented in Tables \ref{table:charged}
and \ref{table:neutral}. As has been noted several times in the literature, the branching ratios in the neutral
case are very sensitive to the values of the inputs and this continues to be the case
when the 27-plet contributions are added. Thus the philosophy of taking one loop ChPT 
predictions as an indicator of the order of magnitude or as providing a limit 
continues in the present case.

\begin{table*}
\begin{center}
\caption{Branching fractions for $K^+ \rightarrow \pi^+l^+l^-$. 
using central values of the inputs from \cite{Cirigliano:2011ny,Bijnens:2004vz}} 
\label{table:charged}
\begin{tabular}{ccc}
\noalign{\smallskip}\hline
 &~~~~~~~$K^+ \rightarrow \pi^+e^+e^-$	&~~~~~~~$K^+ \rightarrow \pi^+\mu^+\mu^-$ \\
\noalign{\smallskip}\hline\noalign{\smallskip} 
octet			&~~~~~~~3.86 $\times 10^{-7}$ 		&~~~~~~~8.49 $\times 10^{-8}$ 	\\\vspace{0.05cm}
octet and 27-plet	&~~~~~~~4.37 $\times 10^{-7}$ 		&~~~~~~~9.30 $\times 10^{-8}$ 	\\
\noalign{\smallskip}\hline\noalign{\smallskip} 
\end{tabular}
\end{center}
\end{table*}

\begin{table*}
\begin{center}
\caption{Branching fractions for the $K_S \rightarrow \pi^0l^+l^-$ using
central values of the inputs from \cite{Cirigliano:2011ny,Bijnens:2004vz}} 
\label{table:neutral}
\begin{tabular}{ccc}
\noalign{\smallskip}\hline
 &~~~~~~~$K_S \rightarrow \pi^0e^+e^-$	&~~~~~~~$K_S \rightarrow \pi^0\mu^+\mu^-$ \\
\noalign{\smallskip}\hline\noalign{\smallskip} 
octet			&~~~~~~~5.43 $\times 10^{-7}$ 		&~~~~~~~1.17 $\times 10^{-7}$ 	\\\vspace{0.05cm}
octet and 27-plet	&~~~~~~~6.81 $\times 10^{-7}$ 		&~~~~~~~1.52 $\times 10^{-7}$ 	\\
\noalign{\smallskip}\hline\noalign{\smallskip} 
\end{tabular}
\end{center}
\end{table*}

In order to address the issue of the extent of unitarity corrections to the octet
part of the amplitude, we have taken the expressions from ref.\cite{D'Ambrosio:1998yj} 
(i.e., amplitudes with the $G_{27}$ contribution missing). In Fig.\ref{fig:fig5},
we plot the quantity $W(z)$ \footnote{Here $W(z) =  G_F^2 M_K^4 V_+(z)$} as 
a function of $z$ and scaled by a factor of $10^{12}$.
We compare our results which account for the full amplitude (both $G_8$ and
$G_{27}$) without unitarity corrections with that of the 
unitarity corrected amplitude from ref.\cite{D'Ambrosio:1998yj}.
We have plotted our results for $G_8$ alone and with the $G_{27}$ included (central values for both) whereas
we have reproduced the curves given in Figs.3 and 4 of ref.\cite{D'Ambrosio:1998yj}
corresponding to different set of the free parameters $a$ and $b$. 
We see that there is a crossover of our curves with the latter for $a=-0.62, b=-0.3$
and $a=0.47, b=1.5$ while the curves corresponding to $a=-0.68, b=0.0$ and $a=0.55, b=1.1$
falls below our curves. The effect of including the $G_{27}$ part is clearly visible form the figure.
It may be noted that the full unitarity corrections to the complete amplitude we have presented needs to
be computed and is a project for the future.

We would like to discuss the implications of our amplitude to the issue of determination of the LEC's.
To the branching ratios of the charged case, the $G_{27}$ contribution is of the order of $\sim 10\%$ which
in turn implies that if the LEC's are to be determined better, the amplitude given in this paper should be
a valuable input. For instance, the $N_{14}^r,N_{15}^r$ determinations in ref. 
\cite{Bijnens:2004ku,Bijnens:2004vz} is done using the branching ratios for 
charged kaons without the $G_{27}$ contribution. A new analysis which would modify these estimates a little 
would be welcome. Furthermore, the LEC's along with the $G_8$ and $G_{27}$ could be fitted to the 
distributions obtained by \cite{Batley:2009pv,Batley:2011zz}.

We summarize here the contents of this paper. Whereas $K \rightarrow \pi l^+l^-$ remains an
important and interesting process studied in one loop $SU(3)$ ChPT and beyond by using unitarity,
the 27-plet contribution to the amplitude given in eqns.(\ref{eq:charged}) and (\ref{eq:neutral})
was not evaluated earlier. Here we present the full 
one loop $SU(3)$ ChPT amplitude which we obtained from the software described in ref. \cite{Unterdorfer:2005au}.
Using recent estimates for $G_8, G_{27}, N_{14}^r$ and $N_{15}^r$, we have established that there is a noticeable 
contribution to the charged decay spectrum as well as to the rate, with the latter at the $\sim 10\%$ level.
The corresponding contribution in the case of the neutral decays and is also of the order of $\sim 10\%$ 
for \cite{Bijnens:2004vz}. We have also performed an analysis of the uncertainties associated with the
dominant octet part in order to assess the impact of the 27-plet contribution to the amplitude.
Since these decays have played a vital role in the determination of the LEC's,
the new amplitude could be used in the future for fits to experimental data.  
It is hard to directly compare our results with that of the recent measurements, which have not used the most 
up to date values of the constants and have not accounted for the $G_{27}$ contributions feeding into the analysis from the piece in the amplitudes
given in eqns.(\ref{eq:charged}) and (\ref{eq:neutral}) either.
An improvement of this full amplitude using unitarity as in \cite{D'Ambrosio:1998yj} could also warrant a full investigation.
It is our belief that a comprehensive analysis of the data including $G_{27}$
effects would be a worthwhile exercise, with some of the constants fixed from other
experiments.  It may give rise to a determination of the $R's$. 
We would like to point out that the 27-plet 
contribution to $K^+ \rightarrow \pi^+ \gamma \gamma$ has been treated
in ref.\cite{D'Ambrosio:1996zx} through unitarity corrections and the complete
expression is given in ref.\cite{Gerard:2005yk} The corresponding contribution 
to $K \rightarrow \pi \pi \gamma$ is given in ref.\cite{Mertens:2011ts}
Thus, our work provides a complement to these efforts.

\vskip0.5cm
{\small {\bf Acknowledgements:} 
We are very grateful to Gerhard Ecker for extensive correspondence on the subject
and to Bastian Kubis, Johan Bijnens and Gauhar Abbas for discussions.}

\end{document}